# Counterdiffusion-in-gel growth of high optical and crystal quality MAPbX$_3$ (MA = CH$_3$NH$_3^+$, X = I$^-$, Br$^-$) lead-halide perovskite single crystals

Nikita I. Selivanov,[1,*] Aleksei O. Murzin,[1] Vsevolod I. Yudin,[1] Yury V. Kapitonov,[1] Alexei V. Emeline[1]

[1]St. Petersburg State University,
Ulyanovskaya 1, St. Petersburg, 198504, Russia,
*Corresponding author: N. I. Selivanov
E-mail address: Selivanov, N. I. (selivanov_chem@mail.ru)

Halide perovskites are promising semiconductor materials for optoelectronics. One of their unique features is the synthesis, that could be performed by the crystallization from the solution. This method is fast, but practically couldn't be optimized due to the strongly non-equilibrium conditions of crystal growth. Another quality limitation is the typical incorporation of organic solvent molecules into crystals when perovskite crystallization is performed from solution with organic solvent.. Here we report an alternative method of perovskite single crystal growth based on the counterdiffusion of reagents in the gel medium. This counterdiffusion-in-gel crystallization (CGC) method is based on the difference of solubility between perovskites and lead halide in the corresponding halide acid, HX. The difference of solubility between perovskites and lead halide leads to the formation of perovskite single crystals of sufficiently large size with an excellent optical quality at constant room temperature. Particularly, here we report the growth of MAPbI$_3$ and MAPbBr$_3$ halide perovskite single crystals (MA = CH$_3$NH$_3^+$) by the counterdiffusion of the organic-solvent-free reagents in the U-tube filled with the silica gel. Their photoluminescence spectra recorded at 4 K demonstrate the excitonic resonances with the full width < 2.5 meV, which proves their excellent optical quality. The proposed method can be used in fabrication of high-quality halide perovskite single crystals for both fundamental research, and for applications where the absence of defects is a critical requirement.

**Introduction**

Halide perovskites, the optical properties of which have been studied in the background for many decades [1], [2], quickly drew attention after being proposed as a light-absorbing material for solar cells [3]. The ease of formation of their thin polycrystalline films and the unique optoelectronic properties have led to the creation of cost-competitive halide perovskite solar cells with an efficiency above 25% [4]. Once in the spotlight, halide perovskites quickly shone with other facets of their uniqueness, revealing that in the form of single crystals they are excellent materials for a variety of optoelectronic and optical applications, such as light emitting diodes [5], lasers [6], gamma- and X-ray detectors [7] and photodetectors [8]. The large exciton binding energy [9] also favors their use in information photonics [10] and polaritonics devices [11]. All these promising applications are united by the need for high quality single crystals of ABX$_3$ halide perovskites, where currently in the vast majority of cases A = Cs$^+$, MA (MA = CH$_3$NH$_3^+$), B = Pb$^{2+}$ and X = I$^-$, Br$^-$.

The main approaches for growing halide perovskite single crystals are based on the crystallization from the solution, including solution temperature-lowering (STL) [12,13], inverse temperature crystallization (ITC) [14-16], and anti-solvent vapor-assisted crystallization (AVC) [17,18] methods at near-room temperatures. For MAPbX$_3$ (X = I, Br) growth, dimethylformamide (DMFA) (ITC, AVC), gamma-butyrolactone (GBL) (ITC, AVC) and corresponding hydrohalic acids (STL, AVC) are used as solvents. The possibility of repeated growth using seed crystals allows an almost unlimited increase in crystal volume, which was demonstrated by several-inch-size single crystals grown by ITC [16]. The next step was to slow down the growth to obtain large crystals of higher quality. Low-temperature-gradient crystallization (LTGC) method made it possible to obtain MAPbBr$_3$ crystals of several-inch size and higher structural and optoelectronic quality from the solution PbBr$_2$ and MABr in DMFA by slow heating (2$^o$C/day) for 20 days [19]. Also worth mentioning is the Bridgman growth method applicable to the growth of fully inorganic CsPbBr$_3$ single crystals from the melt of salts [20]. Although this method makes it possible to obtain large crystals, their cooling to the room temperature leads to the accumulation of strain due to the structural phase transition in CsPbBr$_3$ at 130$^o$C [20].

LTGC approach pushed the method of crystallization from the solution to the limit by optimizing the critical parameter – temperature. Despite this, the method contains inevitable drawbacks associated with the flow of the material from the liquid and the thermodynamics of the process itself: convection currents and turbulence, increased number of crystalline imperfections and incorporated impurities due to elevated growth temperature, defects formation and stress introduction at the crystal cooling stage. Additional problems are a necessity for running the preliminary synthesis of methylammonium halides, complex equipment for smooth ramp of the solution temperature, and utilization of the toxic (DMFA) and psychoactive (GBL) solvents, and their incorporation in crystals. In addition, crystals grown in organic solvents by the ITC method, are susceptible to corrosive effects from solvents remaining on the crystal surface after its extraction from solution, which negatively affects their quality [21,22].

In this study, we propose a modification of the method previously applied only for the growth of single crystals of solids almost insoluble in water, such as PbI$_2$, in silica gel [23-25]. The only example of the growth of 3D halide perovskites in silica gel is the formation of the single crystals of rubidium and cesium tin halides, performed without any characterization of their optical properties [26]. Our successful experience of growing the low-dimensional lead halides (halide perovskitoides) with excellent optical properties [27-29] has led us to the utilization of the proposed method to form single crystals of 3D organic-inorganic halide perovskites.

The proposed counterdiffusion-in-gel crystallization (CGC) method is based on the lower solubility of halide perovskites compared to the lead halides PbX$_2$ in the corresponding halide acid HX (so called reaction gel growth [30]). The silica gel used forms a penetrable porous barrier between reagents salts solutions, that ensures their slow counterdiffusion to the region of the perovskite single crystal formation. Such crystal growth conditions are close to the equilibrium between crystallization and dissolution, and therefore, results in the high quality of formed single crystals.

The proposed method is free from the disadvantages inherent in the method of crystallization from solution: gel medium is free from any convection and turbulence, it provides the three-dimensional support of the crystal being comparable to the crystal growth in microgravity, and ensures spatial separation of growing crystals. The constant room temperature results in fewer defects, and emergence of new nucleation centers is suppressed [30]. Gel-growth method makes it possible to control the nucleation rate, number of crystals and their size by controlling reagents concentrations [24]. The nucleation could be further suppressed and the crystal size could be further increased by the re-growth technique, similar to the ITC method. For this, the gel-grown crystal is placed again into the sol, the sol is converted into the gel, and the crystal continues its growth after adding the reagents. Besides, CCG method does not require a special sophisticated equipment for controlled temperature alteration during the crystal growth and could be carried out at room temperature. In other words, this method is attractively simple in its realization.

We apply our new CGC method to grow MAPbX$_3$ (X = I$^-$, Br$^-$) single crystals by the counterdiffusion of the organic-solvent-free reagents at constant room temperature in the U-shaped vessel (U-tube) filled with the silica gel (Fig. 1, a). The crystals obtained have the high optical and crystal quality, which, together with the above advantages, makes the proposed counterdiffusion-in-gel crystallization method extremely attractive for controllable fabrication of high-quality halide perovskites single crystals.

**Experimental**

*Chemicals*

Lead (II) bromide PbBr$_2$ (98%, Sigma-Aldrich), lead (II) iodide PbI2 (99%, Sigma-Aldrich), hydrobromic acid HBr (40% in H$_2$O, Iodobrom), hydroiodic acid HI (56% in H$_2$O, Iodobrom), hypophosphorous acid H$_3$PO$_2$ (50% in H$_2$O, Acros Organics), methylamine CH$_3$NH$_2$ (38% in H$_2$O, Lenreactiv) were used as received. Silica gel was prepared from sodium metasilicatecrystallohydrate solution Na$_2$SiO$_3$·9H$_2$O with the distilled water as solvent.

To stabilize the hydroiodic acid, hypophosphorous acid was added to it in the 9:1 volume ratio. All solutions and sols, where HI was used as a solvent or reagent, were prepared using this stabilized solution with H$_3$PO$_2$.

*Gel growth*

A silica gel matrix was formed in U-shaped tube (U-tube) (Fig. 1, a). For the silica gel preparation, a required weight of Na$_2$SiO$_3$·9H$_2$O was dissolved in water to obtain a solution with the 0.6 M concentration of sodium metasilicate. This solution was added dropwise and with vigorous stirring (to avoid premature gelation) to the hydrohalic acid with the 1:2 volume ratio. The prepared transparent sol was poured into U-shaped glass tubes with the 15–20 mm inner diameter. A sol-gel conversion took place for 48 hours, and a dense homogeneous silica gel was formed.

*Perovskite single crystal formation*

At the second stage, solutions of PbBr$_2$ in hydrobromic acid with 1 M concentration of and PbI$_2$ in hydroiodic acid with 1.2 M concentration were prepared. Solutions of methylammonium bromide (MABr) with 1 M concentration and methylammonium iodide (MAI) with 1.2 M concentration were prepared by adding the methylamine solution to the corresponding hydrohalic acid. The prepared lead halide solution was poured over the silica gel into one of the limbs of the U-tube. A solution of the corresponding methylammonium halide was poured into another limb.

To eliminate the influence of the temperature factor on the growth and structural quality of crystals, U-tubes with feeding solutions were placed in thermostat at $T = 35°C$. After 2-3 weeks the MAPbX$_3$ single crystals were nucleated in the gel, and continued to grow. Fully grown crystals were mechanically removed from gel, washed in the diluted solution of the corresponding hydrohalic acid, and, after drying in the oven at 40°C, were used for further studies.

*Structural characterization*

Powder X-ray diffraction and diffraction rocking curve studies were carried out with a high-resolution X-ray diffractometer Bruker D8 Discover using a long focus X-ray tube CuKα anode. Reflected X-rays were detected using a solid position sensitive detector LYNXEYE. Measurements were carried out at room temperature.

*Diffuse reflectance spectroscopy*

The diffuse reflectance spectra of powders were obtained using the Cary 5000 UV-Vis-NIR spectrometer equipped with a diffuse reflectance accessory DRA 2500. Barium sulphate was used as a reference sample. Powders were prepared by grinding single crystals in the agate mortar.

*Photoluminescence and reflectivity spectroscopy*

Halide perovskite single crystals were mounted in the closed-loop helium cryostat Montana Instruments and cooled down to the temperature $T = 4$ K. The same 10x Mitutoyo micro-objective lens was used to excite the sample and to collect the outgoing light. Custom-made spectrometer with the CCD array was used to capture spectra with the resolution below 0.1 meV. Normal reflectivity was measured using the halogen lamp. The photoluminescence was excited by the Ti:Sapphire cw-laser tuned well above the band gap (745 nm) in the case of $MAPbI_3$, and by the 405 nm semiconductor laser diode in the case of $MAPbBr_3$.

**Result and discussion**

$MAPbI_3$ and $MAPbBr_3$ single crystals were obtained using the CGC method (Fig. 1a) within a few weeks. Figure 1b,d shows U-tubes with gels during a single crystal growth. It is also possible to get several dozens of crystals in one growth attempt by increasing the reagents concentrations in the feeding solutions (Fig. 1c,e). Insets in Figure 2c,d shows single crystals recovered from the gel. $MAPbI_3$ crystal is black, opaque and has a mirror-like facets. $MAPbBr_3$ crystal is orange-colored and transparent, without foreign inclusions.

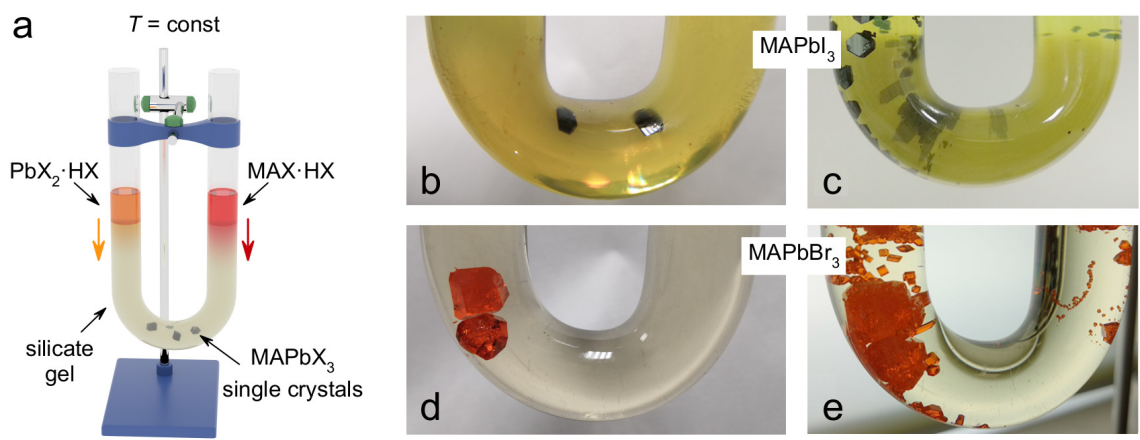

Figure 1. (a) Schematics of the gel growth method. U-tubes with $MAPbI_3$ (b,c) and $MAPbBr_3$ (d,e) crystals grown from gel for low (b,d) and high (c,e) reagents concentrations in the feeding solutions. Tubes diameter is 1.5 cm.

Powder X-ray diffraction (XRD) data for powdered samples (Figure 2a,b) demonstrate a perfect match with the diffractograms calculated based on the single-crystal XRD data [31,32], and prove an absence of non-perovskite phases in the crystals. A high resolution XRD rocking curve method was used to estimate a quality of the crystal structure of the formed perovskite single crystals (Fig. 2c,d). The obtained values of full width at half maximum (FWHM) were 0,021° for $MAPbI_3$ and 0,018° for $MAPbBr_3$, which is similar to the best published data [13,19]. This confirms a very high crystal quality of the grown halide perovskites.

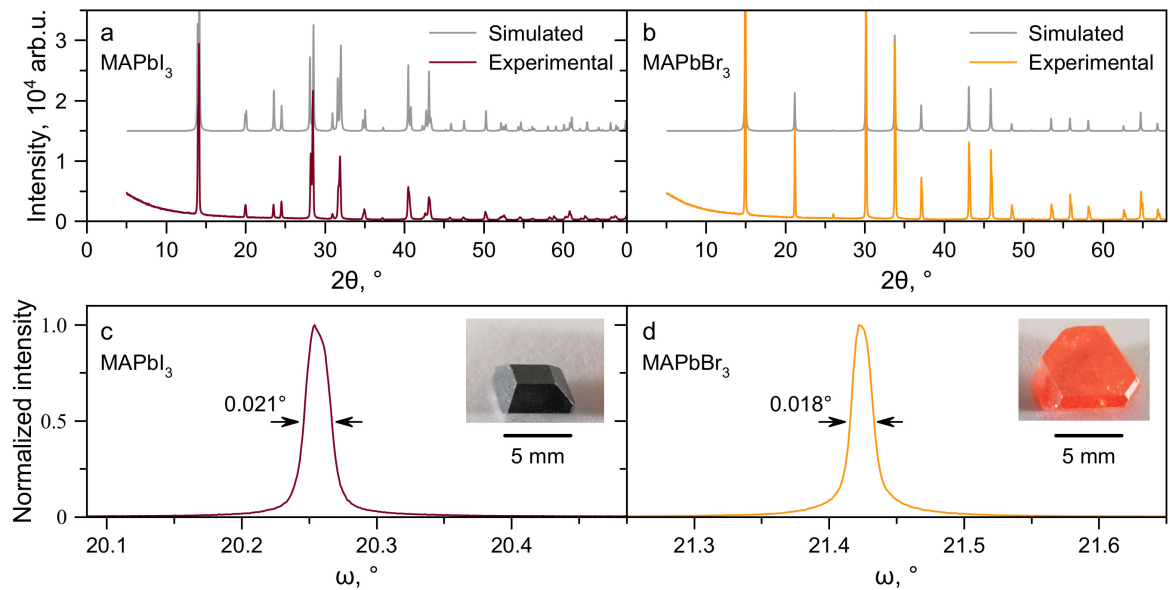

Figure 2. XRD 2θ scans of MAPbI$_3$ (a) and MAPbBr$_3$ (b) powdered crystals and corresponding simulated scans (grey curves). XRD rocking curves of the diffraction peaks of MAPbI$_3$ (c) and MAPbBr$_3$ (d) single crystals. Insets show photos of single crystals extracted from the gel.

Diffusion reflectance spectra (DRS) were measured for powdered crystals at room temperature. Figure 3a shows the Kubelka-Munk function $F(R) = (1 - R)^2 / 2R$, where $R$ is the measured reflectance. For both crystals a sharp fundamental absorption edge is observed, which is typical for the direct band gap semiconductors with vanishingly low density of the defect states within the band gap.

Since in this work we interested in the optical quality of crystals, it is important to choose a suitable method for its evaluation. To exclude the influence of any post-growth fabrication, the all-optical method is preferable. In this work, the analysis of exciton resonance in the reflection and photoluminescence spectra from single crystals at the temperature of liquid helium is chosen as such a method. Since the broadening of the exciton resonance is extremely sensitive to any defects, this method is widely used to analyze the quality of A3B5 semiconductor heterostructures [33], monolayers of transition metal dichalcogenides [34], and halide perovskites [35-37].

Both materials under study are direct-gap semiconductors, which means that excitons (bound states of an electron and a hole) can be observed in them at sufficiently low temperatures. Despite the fact that excitons are hydrogen-like states, the 1s exciton transition plays the major role in the optical properties. Due to the high oscillator strength, the free exciton resonance dominates in the reflection spectra. Exciton emission is also observed in the photoluminescence (PL) spectra. However, in halide perovskites this resonance is usually an exciton bound to a shallow defect [35,36,38]. Measurements at a temperature near 0 K make it possible to completely exclude the influence of temperature fluctuations, and the oscillator strength (and, hence, the radiative broadening) of the bound exciton resonance is negligible. Thus, the width of the bound exciton emission peak near 0 K will be determined by the inhomogeneous broadening and other mechanisms caused by crystal defects. So it can serve as an optical quality factor of the crystal.

The reflectivity and PL spectra from gel-grown MAPbI$_3$ and MAPbBr$_3$ single crystals at 4 K are shown in Fig. 3b and 3c respectively. Both crystals show clear free excitonic transition at 1.637 eV for MAPbI$_3$ and 2.250 eV for MAPbBr$_3$ in the reflectivity [37], and sharp stokes-shifted for 4 meV bound exciton peaks [35] followed by the lower-energy defect-related PL tail [36]. The small full width at half maximum (FWHM) of bound exciton peaks in PL of around 2.5 meV in both cases, together with well-pronounced free exciton resonances in the reflectivity prove the high optical quality of gel-grown halide perovskite single crystals. Their quality is higher than that of solution grown crystals studied in the literature using low-temperature exciton resonance spectroscopy in MAPbI$_3$ [39,40] and MAPbBr$_3$ [37].

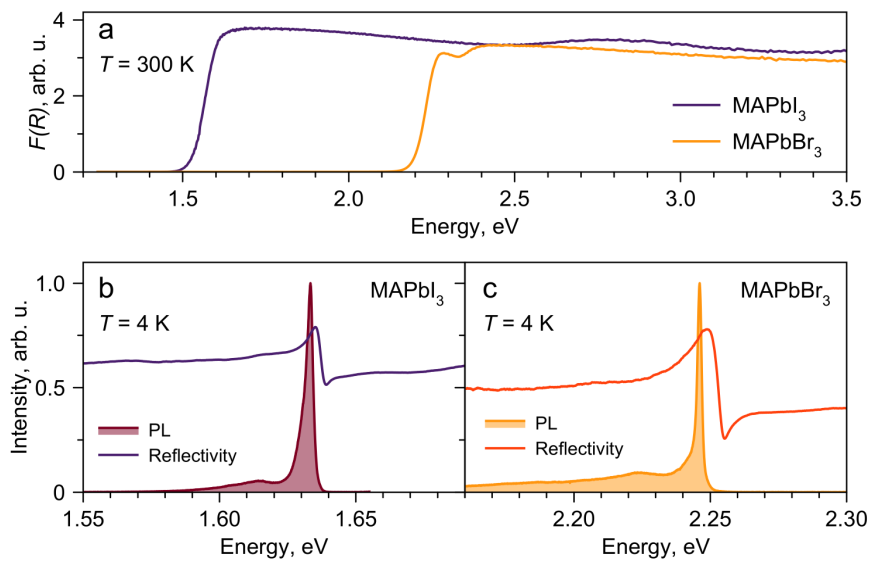

Fig.3. (a) Kubelka-Munk function $F(R)$ for diffuse reflectance spectra of powdered gel-grown perovskite crystals at $T = 300$ K. (b,c) Low-temperature reflectivity and PL spectra from gel-grown MAPbI$_3$ (b) and MAPbBr$_3$ (c) single crystals at $T = 4$ K.

**Conclusions**

We have developed counterdiffusion-in-gel crystallization (CGC) method to obtain high quality MAPbI$_3$ and MAPbBr$_3$ halide perovskites single crystals. The growth in silica gel at constant near-room temperature without organic solvents made it possible to form sufficiently large single crystals with high crystal. Despite the very simple and cheap U-tube growth technique, the crystals obtained demonstrate a high optical quality. Exciton peak FWHM at 4 K for both crystals was on the order of several meV, that is better than previously reported in the literature. The proposed method opens up great prospects both for the synthesis of pure three-dimensional perovskites and a whole galaxy of low-dimensional perovskite-like compounds, and for the growth of doped crystals by introducing the dopant into reactant or directly into the gel. Crystals obtained by this method can satisfy the needs of both fundamental research and specific applications due to their size and quality.

**Acknowledgments**

This study was supported by the Russian Science Foundation (Project No. 19-72-10034). This work was carried out on the equipment of the laboratory "Photoactive Nanocomposite Materials" supported within SPbU program (project ID: 73032813) and SPbU resource center "Nanophotonics". We also are grateful to SPbU resource center "X-Ray Diffraction Studies" for XRD data acquisition and analysis.